\documentclass[showpacs,jkps,floatfix,twocolumn]{revtex4}
\usepackage{graphicx, times}
\usepackage{amsmath}
\newcommand{\ba}{\begin{eqnarray}}
\newcommand{\be}{\begin{equation}}
\newcommand{\ea}{\end{eqnarray}}
\newcommand{\ee}{\end{equation}}

\begin{document}
\title{Directional current-current correlation functions in a two-species hard-core bosons in one dimensional finite lattice: An exact diagonalization study} 
\author{Ji-Woo Lee}
\email{jwlee@mju.ac.kr}
\thanks{Fax: +82-31-335-7248} 
\affiliation{Department of Physics, Myongji University, Yongin 17052, Korea}

\date{\today}

\begin{abstract}
We study a model for two-species hard-core bosons in one dimension.
In this model, the same-species bosons have a hard-core condition at the same site, while different-species bosons are allowed to occupy the same site with a local interaction $U$.
At half-filling, by Jordan-Wigner transformation, the model can be exactly mapped to a fermionic Hubbard model.
Due to this correspondence, the phase transition from superfluid ($U=0$) to Mott insulator ($U>0$) can be explained by simple one-band theory at half-filling.
By using an exact diagonalization method adopting a modified Lanczos method, we obtain the ground states as a function of $U$ for the lattice size upto $L=16$.
We calculate directional current-current correlation functions in this model, which indicate that there are some remaining counter-flow in the Mott insulating region ($U>0$) and co-flow in the charge-density-wave region ($U<0$) for the finite lattices.

\end{abstract}


\maketitle


\section{Introduction}

Investigating quantum many-body systems with strong correlations is one of the most interesting research topics in condensed matter physics.
Also, in atomic physics, trapping quantum particles within an optical lattice enabled researchers to study the collective behavior of quantum particles in a more controllable way.
After the Bose-Einstein condensation of ultra cold atoms is achieved \cite{Wieman1995}, researchers pursued other possible quantum many-body states. 
Greiner {\it et al.} built a system of bosons in a two-dimensional optical lattice and observed a quantum phase transition between Mott-insulator (MI) and superfluid (SF)\cite{Greiner2002}, which has long been studied theoretically by many physicists\cite{Fisher1989}.
Recently, the researchers achieved making artificial Kagome lattice to employ the lattice frustration into quantum many-body systems\cite{S-Kurn2012}.
Since one can build artificial lattices in one, two, or three dimensions at hand, the quantum particles can be used as a simulator of the models of many-body quantum systems with a long history, such as Hubbard model, boson Hubbard model, {\it etc.}

Two-species bosonic systems are also experimentally realized \cite{Mandel} with ${}^{87}$Rb atoms  because alkali atoms may have many internal states possible.
Many exotic phases were reported by numerous authors\cite{Altman2003, Kuklov2003, Isacsson2005, Kuno2013}, which are especially
related to paired bosons with nearest neighbors such as valence-bond states.
This is because bosons have many degrees of interaction between them and there are many possible fillings of interacting bosons.
The interesting phases at zero temperature are the counter-flow and co-flow superfluids.
When the interactions between different species bosons are repulsive and relatively small than the kinetic energy of bosons, overall superfluidity is zero but each species bosons nonzero superfluidity.
Also, when the interactions are attractive and relatively small, overall superfluidity is doubled because the bosons are paired and move together.
But it is very hard to measure directly a quantity which can be used for distinguish whether the bosons are moving in opposite directions or in the same direction.

In this paper, we study a minimal model of two-species bosons in the finite lattices in one dimension at zero temperature. 
For the two species, we denote them as $a$ and $b$ boson.
To restrict the Hilbert space for exact diagonalization, we impose a hard-core condition for each species.
Therefore the possible quantum states for a single site are $|0\rangle,  |a\rangle, |b \rangle, |(ab) \rangle$, where $|0\rangle$ denotes the vacuum state.
Here, $|(ab) \rangle$ state will have the interacting term $U$.
The locally interacting term $U$ between different species bosons at the same site can be either attractive and repulsive.
The model can be regarded as the bosonic version of Hubbard model in one dimension.
Especially, at half filling, the model has three quantum phases at zero temperature: a superfluid ($U=0$), Mott insulator ($U>0$), and paired charge density wave ($U<0$).
When $U$ is positive, the system behaves like an insulator, which is called Mott insulator. The system is also an insulator when $U$ is negative because the paired bosons hinders the movement of other pairs and the phase is called paired charge density wave. 

This paper is organized as follows: 
In the following Section II, we present our model of interacting two-species hard-core boson model with local interaction $U$.
Section III shows the ground-state energies we obtained by using an exact diagonalization method and the derivatives of ground-state energies for the finite lattice system up to $L=16$.
In Section IV, we suggest that we can distinguish the co-flowness  and counter-flowness of bosons with the directional current-current correlation functions.
These functions show that in the Mott insulator, the count-flow correlation is always larger than that of the co-flow correlation.
For charge-density-wave region, the co-flow correlation is always larger than that of the counter-flow correlation and  the difference is exactly the same as that for symmetric value $|U|$ because one can map the one species boson to hole representation and the hole is acting as repulsive boson in the case of $U>0$.
Even though there is no finite-$U$ superfluid-insulator transitions in this model, it is very interesting that the finite-size system exhibit some remaining current-current correlations in the insulating region.
In Section VI, we discuss our results and future research topic.

\section{Model}

The model we study in this work is a model for two-species bosons which are distinguished by the indices $a$ and $b$ in one-dimensional lattice with periodic boundary condition at half-filling.
Therefore, the number of $a$ ($b$) bosons is the half the lattice size.
Each species has a hard-core condition so that the possible number of each boson species in each site is restricted to 0 or 1. 
If there is no interaction between $a$ and $b$, the system can be regarded as two separate bosonic systems.
We introduce a local interaction $U$ between $a$ and $b$ boson. 
In this sense, this model is a minimal interacting model of a two-species bosonic system.

The Hamiltonian is written as follows:
\begin{equation}
H = -t \sum_{
\langle i, j \rangle, \sigma=a, b } ( b^+_{i \sigma} b_{j\sigma}  + H.c. )
+ U \sum_{i} n_{ia } n_{ib},
\label{Hamiltonianeq}
\end{equation}
where
$b^+_{i \sigma} (b)$ is the creation (destruction) operator at site $i$ with a species index of $\sigma$($= a$ or $b$), 
and $n_{i \sigma} \equiv b^+_{i\sigma} b_{i \sigma}$.
Here $t$ is the hopping energy between nearest neighbors, and $U$ is on-site Coulomb interaction energy between $a$ and $b$ when they occupy the same site $i$.
In the following, the energies are scaled as we set $t=1$.

The only changing parameter in this model is the local interaction between two species bosons.
It is interesting to note that this model has a correspondenc to fermionic Hubbard model if one regards boson species as electron spins, although this model  has a fundamental difference because the particles have a different symmetry.
The fermionic Hubbard model becomes Mott insulator as we turn on $U>0$. 
The metallic phase is possible only when $U=0$.
Similarly, the bosons are superfluid at $U=0$ and it becomes
Mott insulator when $U>0$.
Also, for two-species softcore bosons, it was reported that the bosons have two kind of superfluid, which are co-flow and counter-flow superfluid in the non half-filling\cite{Isacsson2005}.
For the perfect counter-flow, the overall superfluidity will disappear but for the imperfect counter-flow, the system will be a weak superfluid.
For attractive bosons, they will form a co-flow superfluid.
When this work started, we were interested how to quantify counter-flowness or co-flowness in bosonic systems. 
We started from measuring of some correlations of bosons from the half-filling and we found some interesting behaviors of nonlocal correlations of currents in this model.

\section{Ground-state energy and Double occupancy}

The system we study is at half filling. 
The number of $a$-boson and that of $b$-boson are both $L/2$ where $L$ is the system size and takes even number for convenience.
We adopted a modified Lanczos method\cite{Gagliano1986} to calculate the ground state and the ground state energy up to $L=16$. 
Because there are 4 possible state in each site, the dimension  of Hilbert space  for $L=16$ is
$({}_{16} {\rm C}_8 ) ^2 = (12,870)^2 = 165,636,900$.
Obtaining the diagonal elements of the Hamiltonian is straightforward.
But the tricky part is obtaining the off-diagonal elements by $t$ term in Eq. (1).
There exists an easy way to calculate $t$ term due to the fact that one species does not interfere with the configuration of the other species.
We adopted a multi-core parallel programming method by OPENMP to obtain the ground state\cite{OPENMP}.
We used computing resources of 39GB memory of 12-thread Xeon 2.4GHz CPU and it took 1 day to obtain a single point for $L=16$.

\begin{figure}[t]
\includegraphics[width=8cm]{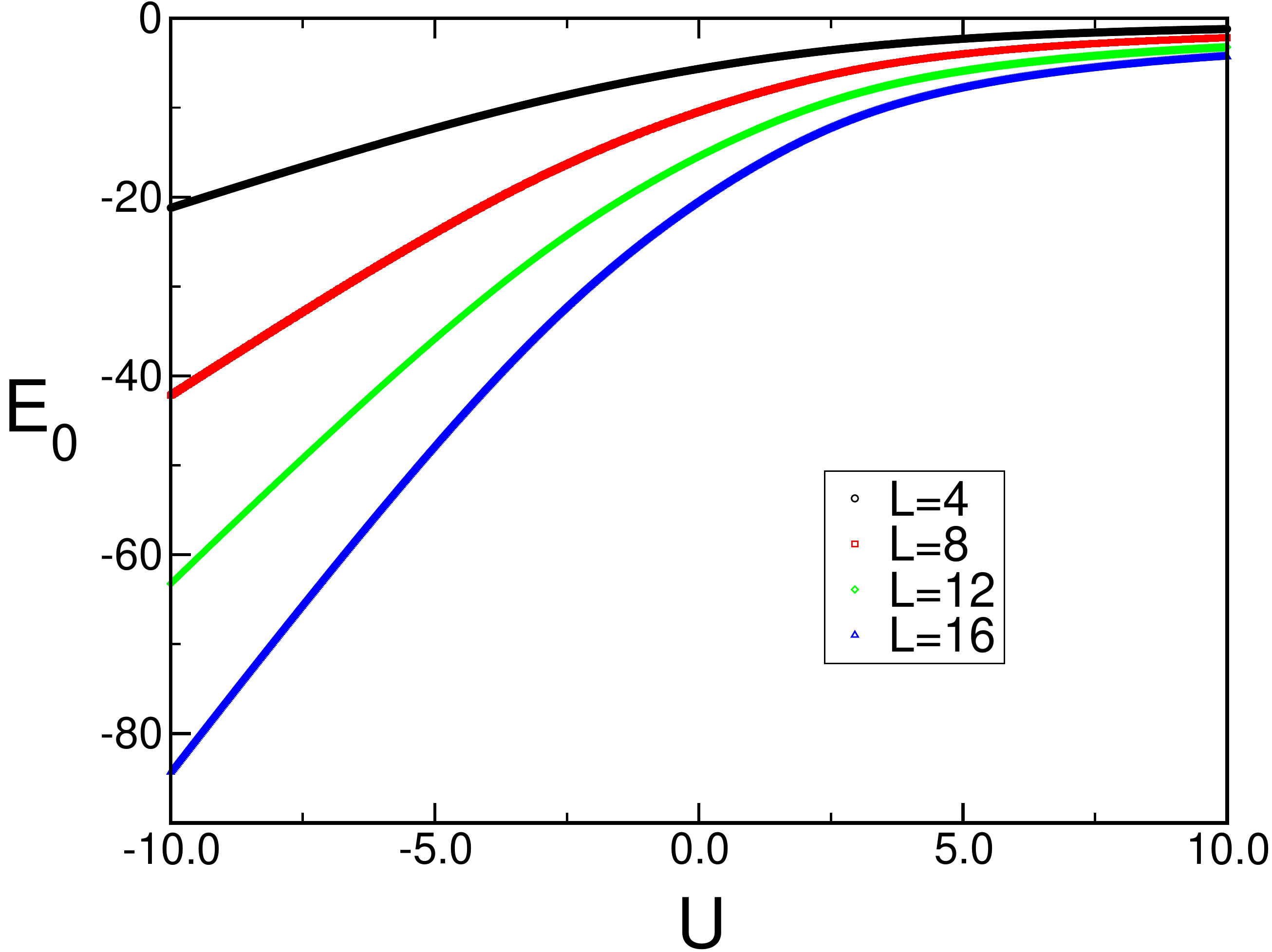}
\vspace*{0.0in}
\caption{(color online)
Ground-state energies of two-species hard-core Hamiltonian (Eq. (\ref{Hamiltonianeq})) as a function of $U$ for $L = 4, 8, 12$, and $16$ by using exact diagonalizations.
The step in $U$ is 0.02 so that the data points are dense enough to look like a line. 
\label{fig-ge}}
\end{figure}

In Figure 1, we show the ground-state energy $E_0$ as a function of $U$ for the lattice sizes, $L=4, 8, 12$, and $16$.
For the negative $U$, the two boson species tends to stick together at a local site because of the attraction between different species.
As one can notice, as $U \rightarrow \infty$, the energy goes to 0 since each bosons will occupy every other sites alternately. 
Because the bosons are frozen at local sites, the energy becomes zero even though there exist hopping energies.

Figure 2 shows the ground-state energies per site ($\epsilon_0 \equiv  E_0/L$)  as a function of $U$.
Interestingly, the curves are almost collapsed for $L=8, 12$, and $16$.
This results show that our exact diagonalization calculation is reliable.
Especially, for $U=0$, we know that exact number by the formula,
\begin{equation}
2 \times\Bigg( \sum_{i= \pm 2, \pm 4 , \cdots} -2 \cos \frac{ i \pi}{L}  \Bigg)
\end{equation}
so that $\epsilon_0(L=16) = -1.281457724$.
The value of our calculation is $-1.28145767145$
for $L=16$.
We truncated the exact diagonalization at the accuracy of $O(10^{-7})$.
\begin{figure}[t]
\includegraphics[width=8cm]{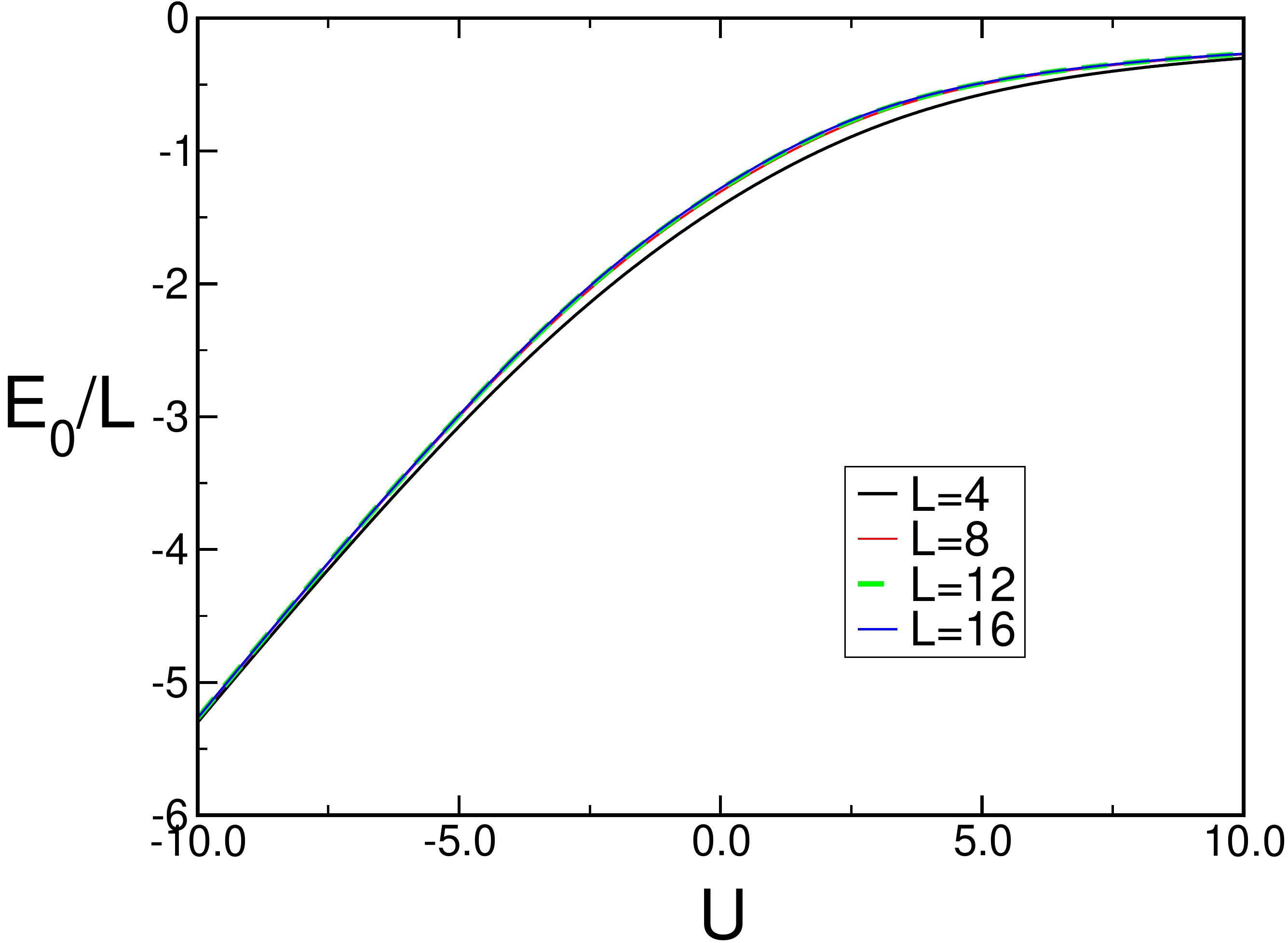}
\vspace*{0.0in}
\caption{(color online)
The ground-state energies per site ($\epsilon_0 \equiv E_0/L $) as a function of $U$ for L=4, 8, 12, and 16.
The step for $U$ is 0.02.
The curves are almost collapsed to a single line as we increase the system size.
\label{fig-gepersite}}
\end{figure}


We can define a double occupancy as
\begin{equation}
\langle \frac{1}{L} \sum_i  n_{ia} n_{ib}  \rangle.
\end{equation}
We obtained the double occupancy by differentiating the ground-state energy with respect to 
$U$.  
Figure 3 shows the double occupancy as a function of $U$.
The double occupancy falls from 0.5 to 0 as we increase $U$.
In the middle of $U$, the double occupancy shows a linear behavior and we believe
that this region corresponds to a superfluid phase.
\begin{figure}[b]
\includegraphics[width=8cm]{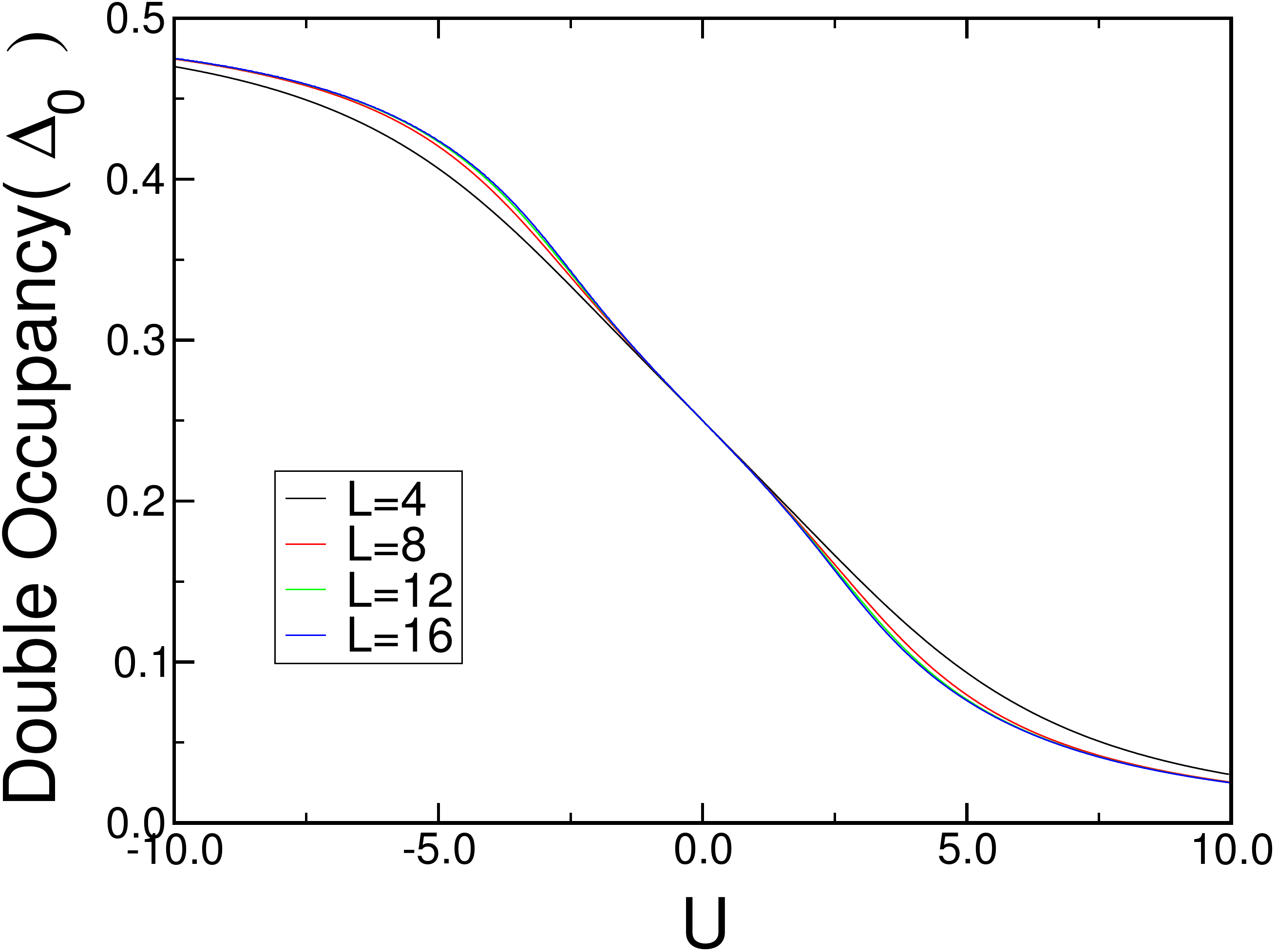}
\vspace*{0in}
\caption{(color online)
Double occupancy of two-species bosons.
When $U$ is negatively large, the half of the lattice size is filled with two bosons so that
the double occupancy becomes 0.5.
When $U$ is positively large, the bosons are located alternately so that the double occupancy
becomes 0.
In the middle around $U=0$, the double occupancy is a linear function and collapse to a single curve.
\label{rengap}}
\end{figure}

 \begin{figure}[t]
\includegraphics[width=8cm]{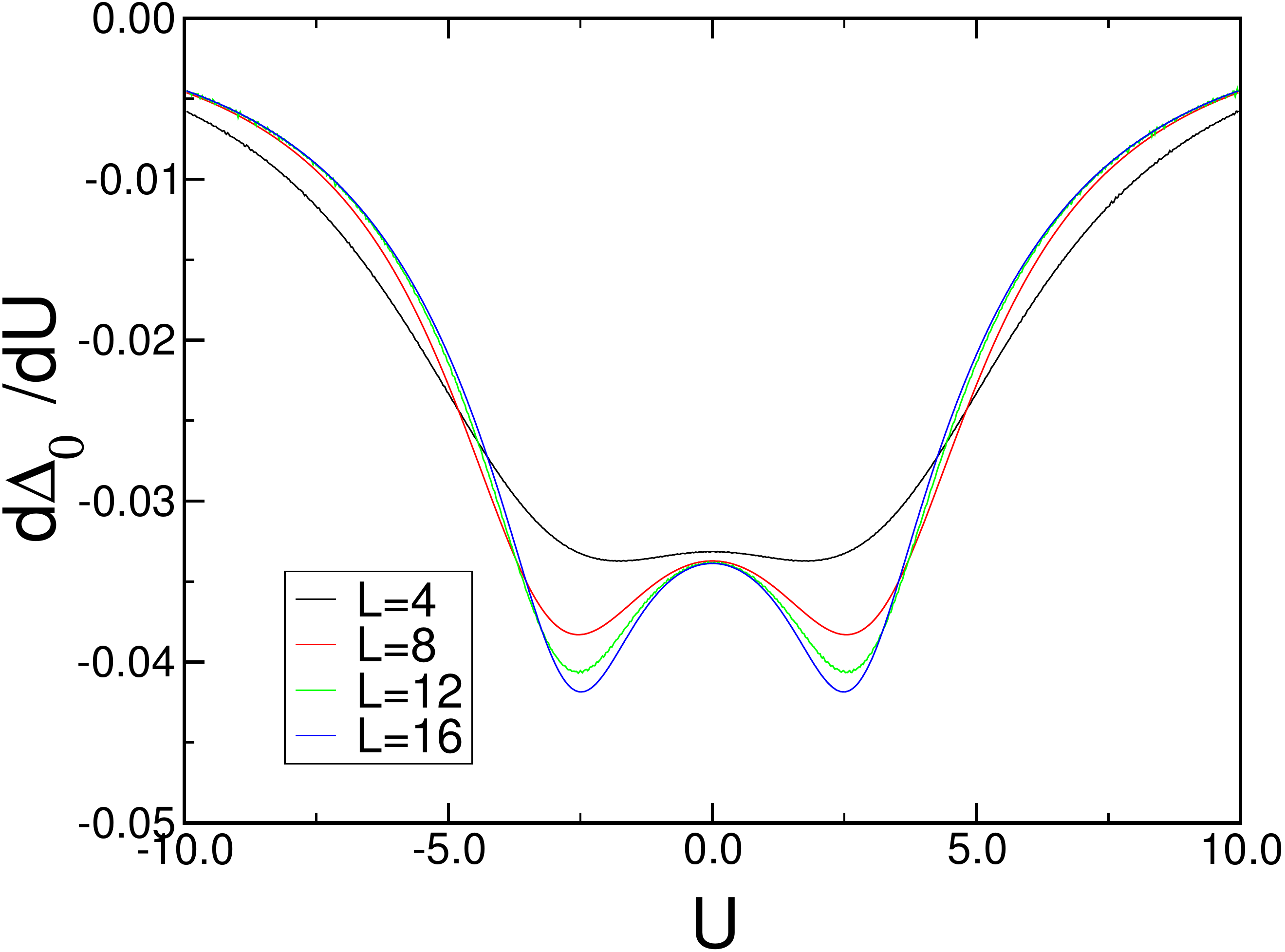}
\vspace*{-0.1in}
\caption{(color online)
Derivatives of double occupancy as a function of $U$.
As the system size increases, the local minima decreases but the
local maximum at $U=0$ remains near the value of  $-0.0339$.
\label{derdo}
}
\end{figure}
Figure 4 shows the derivatives of double occupancy as a function of $U$.
Interestingly, the graph show a two local minima and a local maximum.
As one can see clearly, the $|U|$ values of local minima decreases as we increase the system size.
For $L=16$, the minima occur at $U \approx \pm 2.50$.
This results are consistent with the infinite-size Bethe Ansatz energies\cite{Korepin}.

\section{Directional current-current correlation functions}

As we explained in Section I, it is not an easy task to distinguish the co-flowness and the counter-flowness.
In this Section, we develop  directional current-current correlation functions that will distinguish the co-flowness and counter-flowness. 
We define co-flow current-current correlation function as
\be
 C_{RR}=  \langle \frac{1}{L^2} \sum_{jk} j_R(i)j_R(k) \rangle  
\ee
or
\be
 C_{LL}=   \langle \frac{1}{L^2} \sum_{jk} j_L(i)j_L(k) \rangle
\ee
where $j_R(i) = b^\dagger_{i+1}  b_i$ which is right-moving current and $j_L(i) =  b^{\dagger}_{i}b_{i+1}$ which is left-moving current.
By  using the symmetry of the system, $C_{RR} = C_{LL}$ and we define $C_{co} \equiv C_{RR}$.
The counter-flow current-current correlation function is defined as
\be
 C_{RL} = \langle \frac{1}{L^2} \sum_{jk} j_R(i) j_L(k) \rangle
\ee
or
\be  
 C_{LR} = \langle \frac{1}{L^2} \sum_{jk} j_L(i) j_R(k) \rangle
 \ee
and  $C_{counter} \equiv C_{RL} = C_{LR}$. 
Here, we note that $\langle \cdots \rangle$ denotes the expectation value with respect to the ground state.
 In general, the current-current correlation function is defined as the correlation between the net current at $i$ and $j$, we found that this is exactly zero because the left and right moving currents are both counted.

In Figure 5, we show the $C_{RR}$ as a function of $U$ for $L=4, 8, 12$, and $16$. 
For $U>0$, the co-flow correlation function goes to 0 as $U$ increases because
$C_{RR} (-U)  = C_{RL}( U) $ by the symmetry.
In Figure 6, we show the $C_{RL}$ as a function of $U$.
The counter-flow correlation function seems to decrease very slowly as $U$ increases.
We believe that the counter-flow correlation goes to 0 when $U \rightarrow \infty$.

If we subtract $C_{RR}$ from $C_{RL}$ ($C_{counter} - C_{co}$), we obtain Figure 7, which clearly shows for $U>0$, the counter-flow correlation function is always larger than co-flow correlation function and for $U<0$, the co-flow correlation function is always larger than the counter-flow correlation function.
With these new quantity that can distinguish co-flow and counter-flow superfluid, the overall superfluidity can be divided correctly with the counter-flow superfluid when $U>0$ and the co-flow superfluid when $U<0$.

 \begin{figure}[t]
\includegraphics[width=8cm]{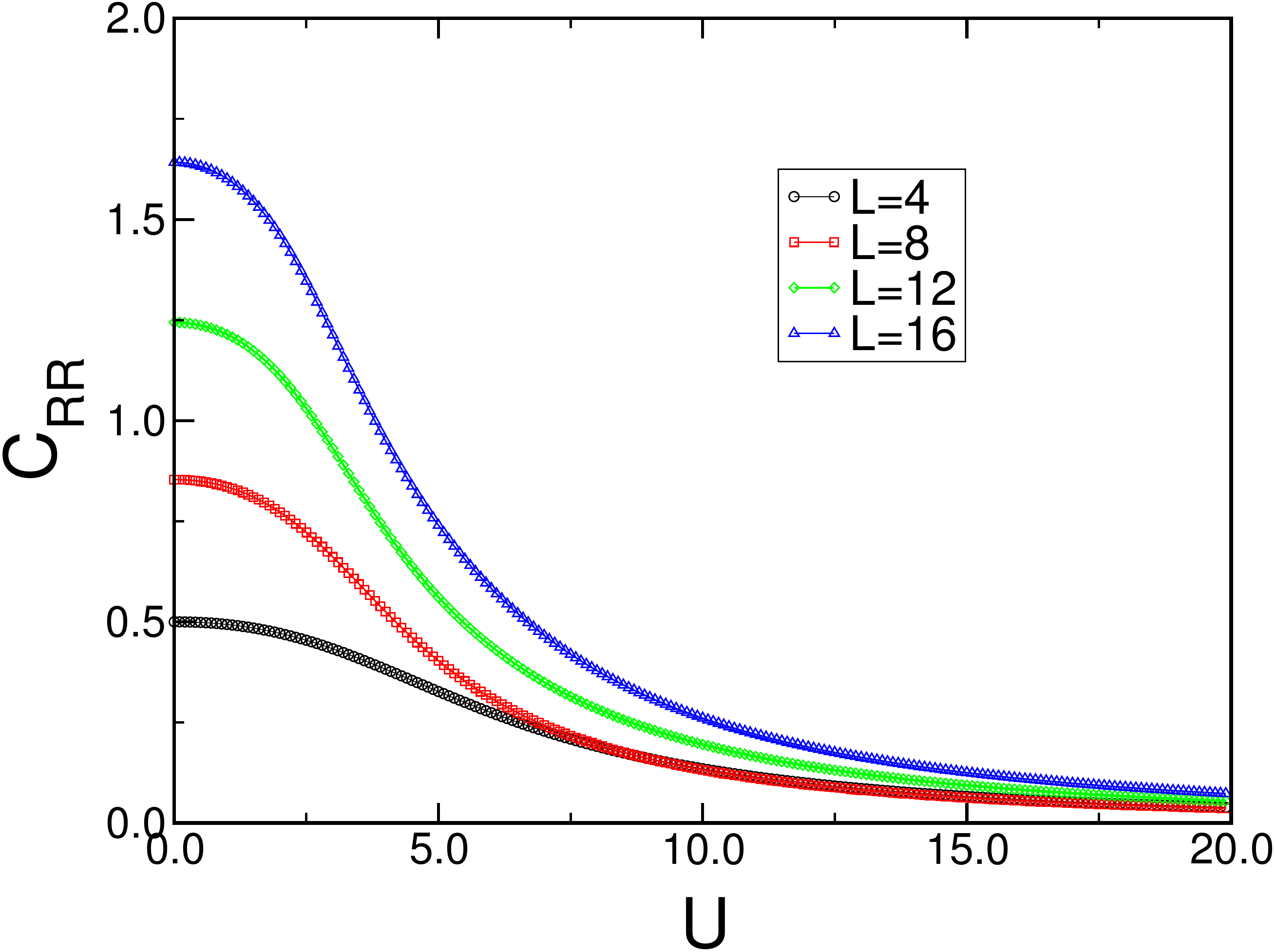}
\vspace*{-0.1in}
\caption{(color online)
Co-flow current-current correlation functions ($C_{co} = C_{RL}$)
as a function of $U$ for L=4, 8, 12, and 16. 
The step of $U$ is 0.1.
We chose sparse data points from the ground states we obtained with the step of 0.02 in $U$ because the calculation time is large when $L=16$.
For $U>0$, the co-flow current-current correlation function goes to 0 as we increase $U$.
\label{rengap}}
\end{figure}

 \begin{figure}[t]
\includegraphics[width=8cm]{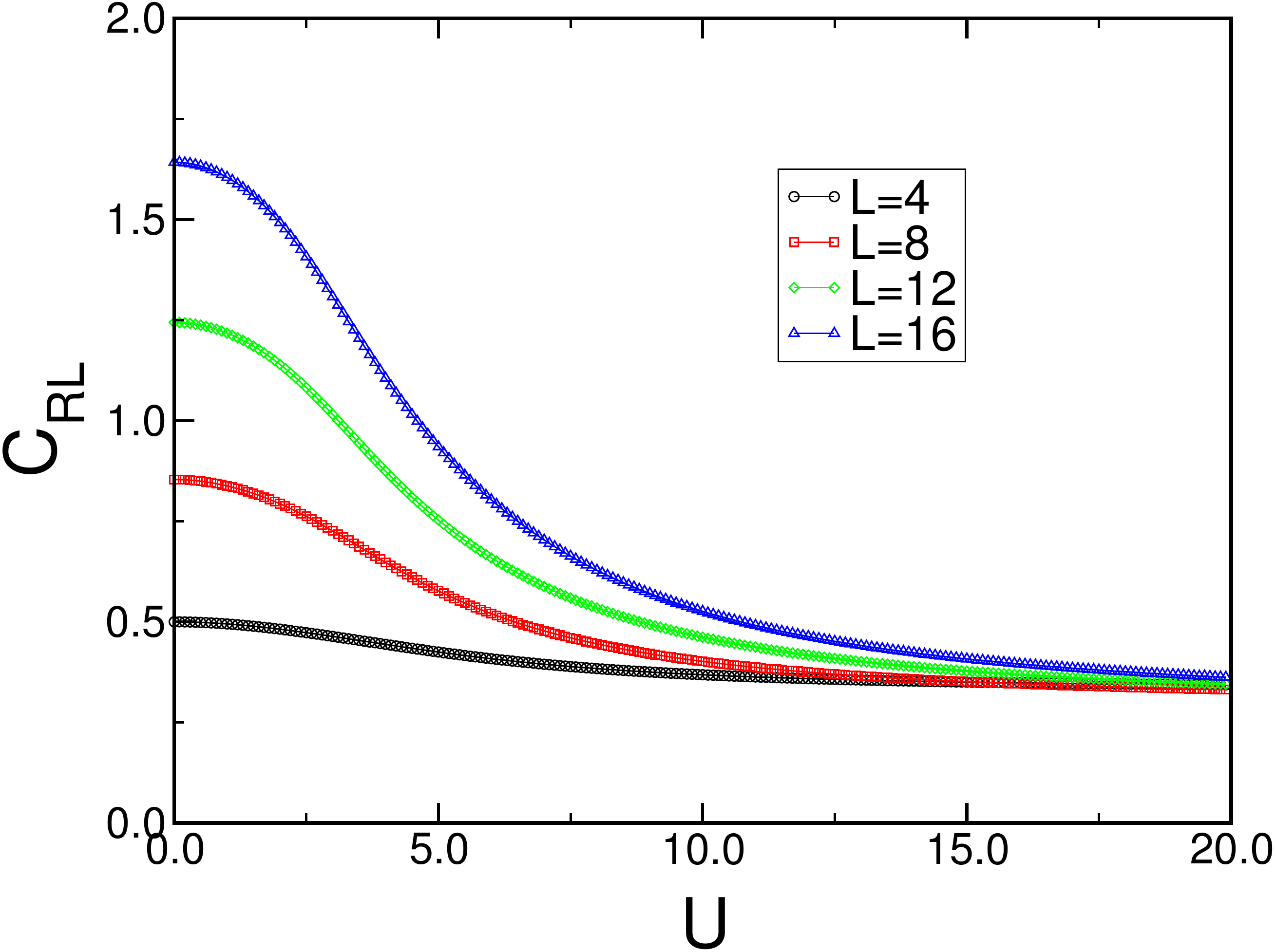}
\vspace*{-0.1in}
\caption{(color online)
Counter-flow current-current correlation functions ($C_{counter} = C_{RR} $ )
as a function of $U$ for L=4, 8, 12, and 16. 
The step of $U$ is 0.1.
We chose sparse data points from the ground states we obtained with the step of 0.02 in $U$ because the calculation time is
large when $L=16$.
For $U>0$, the counter-flow current-current correlation function decreases very slowly as
we increase $U$.
\label{corrRL}}
\end{figure}

 \begin{figure}[b]
\includegraphics[width=8cm]{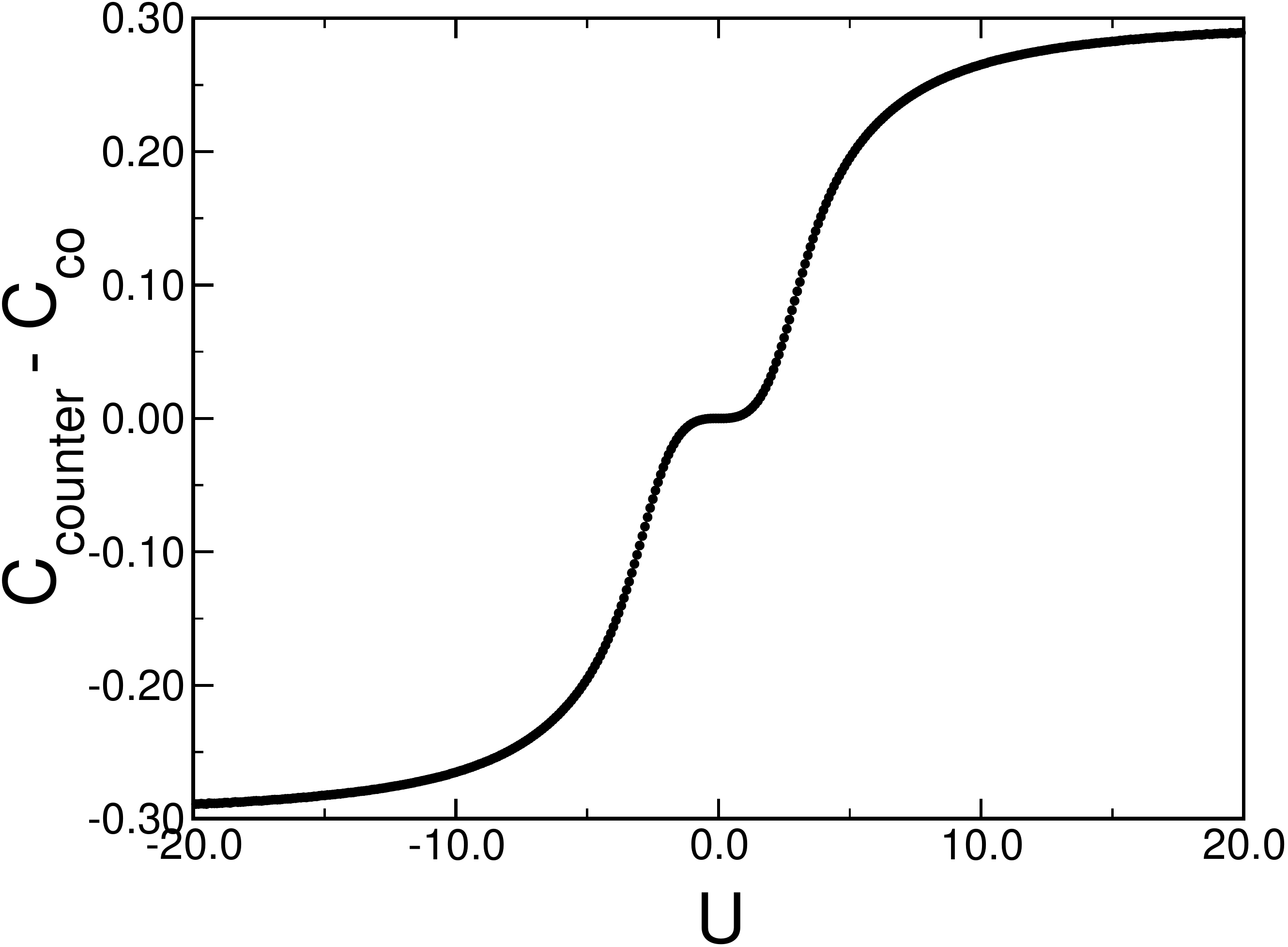}
\vspace*{-0.1in}
\caption{(color online)
Difference of $C_{counter}$ and $C_{co}$ for $ -10 < U <10$ for $L=16$.
The positive value indicates that the counter-flowness is larger than co-flowness 
and the negative value indicates that the co-flowness is larger than counter-flowness.
The correlation remains strong in both insulating phases but we believe
that this correlation goes to 0 as we increase $U$ to $\pm\infty$.
\label{Counter-Co}}
\end{figure}

\section{Summary}

In this paper, 
we studied a model for two-species hard-core bosons in one dimension.
The ground state of this model is obtained with the exact diagonalization method for $L=4, 8, 12$ and $16$.
We found a reasonable behavior of double occupancies.
With the double occupancy, we found that the system from charge density wave insulator to a superfluid (U=0) in the negative $U$ region.
The charge density wave insulator when $U$ is negative has both $a$ and $b$ bosons at a local site, and the ground state is 
\begin{equation}
|\Psi(U<0) \rangle =  | (ab)  \quad 0 \quad (ab) \quad 0  \cdots \rangle.
\end{equation}
For the positively $U$, the system is a Mott insulator in which the $a$ and $b$ bosons occupy alternately, such as
\begin{equation}
|\Psi(U>0) \rangle = | a \quad b \quad  a \quad b \cdots \rangle.
\end{equation}
With exact diagnolization of the model, we found that one can distinguish co-flowness and counter-flowness in the model  with the directional current-current correlation functions.
The counter-flow current-current correlation function is larger than co-flow correlation function for $U>0$ and the co-flow correlation function is larger than the counter-flow correlation function for $U<0$.
It is interesting that these correlations survive deep into the insulating region for the finite-size system.
We believe that these findings can be used in the optical lattice experiments to distinguish the co-flow and counter-flow of bosonic systems. 
We are also working to measure these correlation functions for the unmatched fillings of $a$ and $b$ bosons and we are looking forward to obtaining a supefluid phase with either counter-flowness and co-flowness in those fillings.

\acknowledgments
This research was supported by 2016 Research Fund of Myongji University. 
JWL thanks the Korea Institute for Advanced Study (KIAS) for its hospitality under the Associate Member program.
JWL also thanks M. C. Cha, M. H. Chung, T. Xiang for their stimulating discussions.

\newpage

\end{document}